\documentclass[aip,apl,twocolumn,reprint]{revtex4-2}
\usepackage{graphicx}
\usepackage{dcolumn}
\usepackage{bm}
\usepackage[utf8]{inputenc}
\usepackage[T1]{fontenc}
\usepackage{mathptmx}
\usepackage{etoolbox}
\usepackage{epstopdf}
\usepackage{hyperref}
\hypersetup{hypertex=true,
	colorlinks=true,
	linkcolor=blue,
	anchorcolor=blue,
	citecolor=blue,
	urlcolor=blue}
\draft 

\usepackage{mathtools}
\usepackage[mathlines]{lineno}
\usepackage{environ}	

\newcommand{\tcr}[1]{\textcolor{black}{#1}}

\begin{document}
	
	
\title{Enhanced torque efficiency in ferromagnetic multilayers by introducing naturally oxidized Cu}
	
	
	
\author{Kun Zheng}
\affiliation{Key Laboratory of Polar Materials and Devices (MOE), School of Physics and Electronic Science, East China Normal University, Shanghai 200241, China}

\author{Cuimei Cao}
\affiliation{School of Integrated Circuits $\&$ Wuhan National Laboratory for Optoelectronics, Huazhong University of Science and Technology, Wuhan 430074, China}
	
\author{Yingying Lu}
\affiliation{Key Laboratory of Polar Materials and Devices (MOE), School of Physics and Electronic Science, East China Normal University, Shanghai 200241, China}
	
\author{Jing Meng}
\affiliation{Key Laboratory of Polar Materials and Devices (MOE), School of Physics and Electronic Science, East China Normal University, Shanghai 200241, China}

\author{Junpeng Pan}
\affiliation{Key Laboratory of Polar Materials and Devices (MOE), School of Physics and Electronic Science, East China Normal University, Shanghai 200241, China}

\author{Zhenjie Zhao}
\affiliation{School of Physics and Electronic Science, East China Normal University, Shanghai 200241, China}

\author{Yang Xu}
\affiliation{Key Laboratory of Polar Materials and Devices (MOE), School of Physics and Electronic Science, East China Normal University, Shanghai 200241, China}

\author{Tian Shang}
\thanks{Authors to whom correspondence should be addressed: tshang@phy.ecnu.edu.cn}
\affiliation{Key Laboratory of Polar Materials and Devices (MOE), School of Physics and Electronic Science, East China Normal University, Shanghai 200241, China}

\author{Qingfeng Zhan}
\email[]{qfzhan@phy.ecnu.edu.cn}
\affiliation{Key Laboratory of Polar Materials and Devices (MOE), School of Physics and Electronic Science, East China Normal University, Shanghai 200241, China}
	
	
	
\date{\today}
	
\begin{abstract}
Spin–orbit torque (SOT) in the heavy elements with a large spin–orbit coupling (SOC) has been frequently used to manipulate the magnetic states
in spintronic devices. Recent theoretical works have predicted that the surface oxidized light elements with a negligible SOC can yield a sizable orbital torque (OT), which plays an important role in switching the magnetization. Here, we report anomalous-Hall-resistance and harmonic-Hall-voltage measurements on perpendicularly magnetized Ta/Cu/[Ni/Co]$_{5}$/Cu–CuO$_{x}$ multilayers. Both torque efficiency and spin-Hall angle of these multilayers are largely enhanced by introducing a naturally oxidized Cu-CuO$_{x}$ layer,  where the SOC is negligible. Such an enhancement is mainly due to the collaborative driven of the SOT from the Ta layer and the OT from the Cu/CuO$_{x}$ interface, and can be tuned by controlling the thickness of Cu-CuO$_{x}$ layer. Compared to the Cu–CuO$_{x}$-free multilayers, the maximum torque efficiency and spin-Hall angle were enhanced by a factor of ten, larger than most of the reported values in the other heterostructures.
\end{abstract}
	
	
\maketitle 

Spin-charge interconversion via the spin-Hall effect (SHE) and/or via the inverse-spin-Hall effect (ISHE) has been extensively investigated,\cite{manchon_current-induced_2019,sinova_spin_2015,manchon_current-induced_2019,nakayama_rashba-edelstein_2016,edelstein_spin_1990} 
both of which strongly depend on the strength of spin-orbit coupling (SOC) of the materials used. Spin-orbit torque (SOT) provides efficient and versatile ways to control the magnetic
states and dynamics in various families of materials, and thus, plays a significant role in the spin-based logic devices and nonvolatile memories.\cite{fan_quantifying_2014,miron_perpendicular_2011,liu_spin-torque_2012,manchon_current-induced_2019,sinova_spin_2015} 
Heavy metals (HMs) are generally considered as the nonmagnetic SOT generator due to their large SOC.\cite{woo_enhanced_2014,wang_enhancement_2019,liu_symmetry-dependent_2021}
Very recently, topological insulators and 2D van der Waals materials with a large SOC have been found to generate SOT with a relatively high efficiency.\cite{mellnik_spin-transfer_2014,wang_room_2017,shao_strong_2016,liang_spinorbit_2020}
To data, numerous studies have focused on 
improving the spin-charge interconversion  efficiency by utilizing new materials with a large SOC. 
While the materials with a negligible SOC are generally believed to play minority role in SOT-related phenomena or devices.\cite{sinova_spin_2015}
 
Recently, the orbit torque (OT) has been frequently investigated in the ferromagnetic metal (FM)/$T$ bilayers, where $T$ represents the non-magnetic Ti, Cr and V 3$d$ transition metals.\cite{choi_observation_2023,lee_efficient_2021,xie_efficient_2023,wang_large_2017}  Since the SHE is known to be negligible in these 3$d$ metals, the orbit current is proposed as the mechanism. 
\tcr{The SOT can be interpreted by the mutual precession between the orbital angular momentum and the spin, and thus can be expressed as $\boldsymbol{T}_\mathrm{SO} \approx \xi (\boldsymbol{L} \times \boldsymbol{\sigma})$, where $\xi$ is the strength of the SOC, $\boldsymbol{L}$ is orbital angular momentum and $\boldsymbol{\sigma}$ is the vector of the spin.\cite{go_theory_2020,park_orbital_2012,ho_tunable_2019,hsu_inherent_2018}}
Thus, the orbit current is converted into the spin current in an adjacent FM layer through the SOC, and then, exerts a torque on the local magnetic moments.\cite{choi_observation_2023,lee_efficient_2021,xie_efficient_2023,wang_large_2017}
The oxidized light elements can significantly influence the OT related phenomena, as well as the effective spin-Hall angle, the latter 
determines the conversion efficiency between  charge- and spin currents.\cite{qiu_spinorbit-torque_2015,an_spintorque_2016,kageyama_spin-orbit_2019,okano_nonreciprocal_2019,gao_intrinsic_2018,zheng_disorder_2020,tazaki_current-induced_2020,kim_nontrivial_2021}  
For instance, a large OT can be generated at the interface between  nonmagnetic metal (NM) Cu and oxidized CuO$_x$ in the FM/Cu-CuO$_x$ and the FM/HM/Cu-CuO$_x$ heterostructures.\cite{tazaki_current-induced_2020,xiao_enhancement_2022,ding_harnessing_2020}
In most of these heterostructures, the magnetic moments in the FM layer are aligned within the film plane, which is not preferred for the high-density storage and logical devices.\cite{tazaki_current-induced_2020,kim_nontrivial_2021}
While in the Tm$_3$Fe$_5$O$_{12}$(TmIG)/Pt/Cu-CuO$_{x}$ heterostructure, though the TmIG layer is perpendicularly magnetized, i.e., with the out-of-plane being the easy-axis, the enhancement of torque efficiency due to the Cu-CuO$_{x}$ capping layer is modest.\cite{ding_harnessing_2020} 
In the case of Pt/Co/Cu-CuO$_x$ heterostructure, the critical current for switching the magnetization is comparable to other similar heterostructures without Cu-CuO$_x$ layer.\cite{xiao_enhancement_2022,liu_current-induced_2012,p_spin_2017,xie_situ_2020}
Therefore, combining SOT and OT to enhance torque efﬁciency is expected to be an effective method to further improve the switching efﬁciency in perpendicularly magnetized FM.

\begin{figure}[!htp]
\centering
\includegraphics[width = 0.48\textwidth]{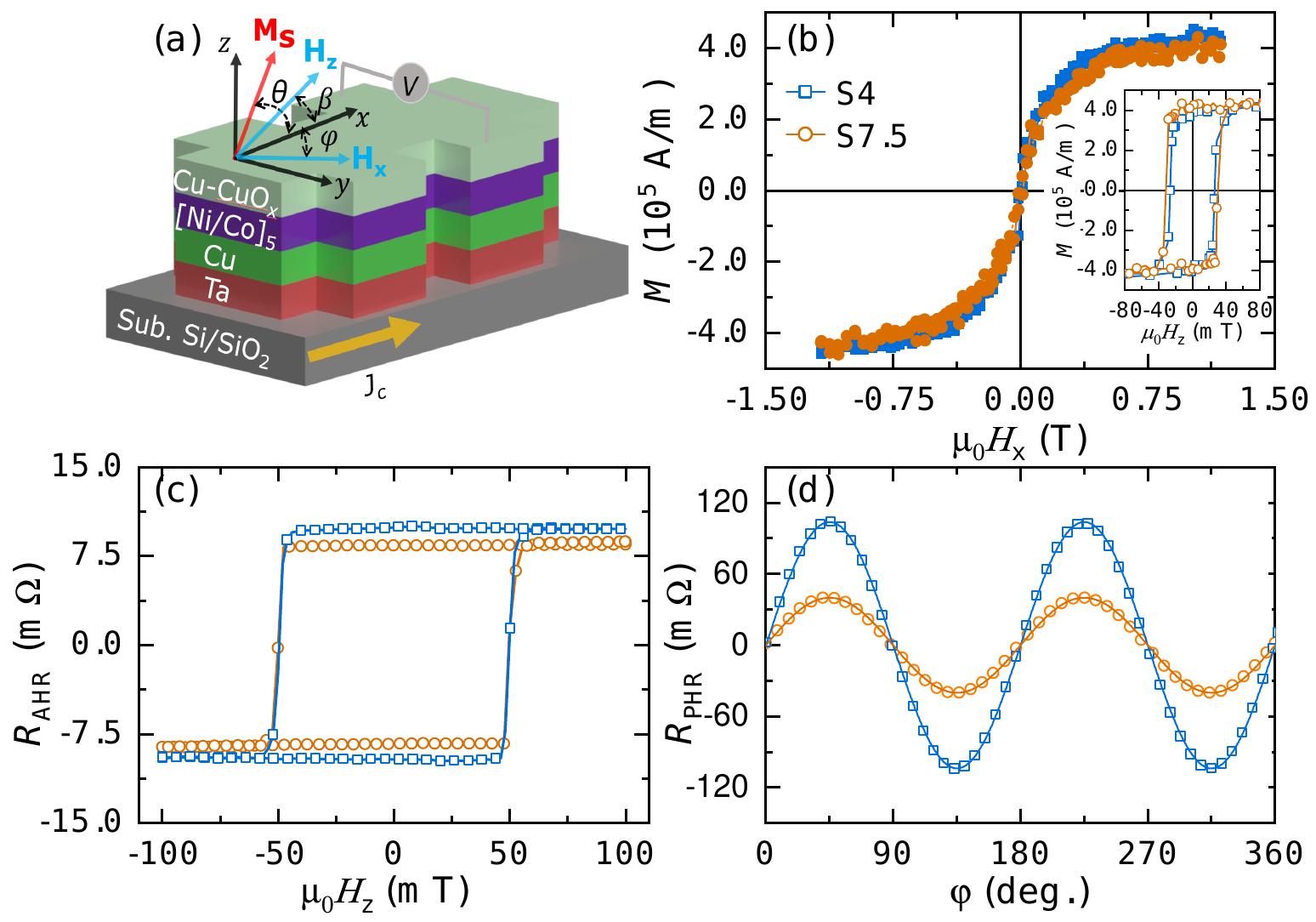}
\vspace{-3ex}%
\caption{\label{fig:fig1}(a) Schematic illustration of Ta/Cu/[Ni/Co]$_5$/Cu-CuO$_{x}$ heterostructures patterned into a Hall-bar geometry. $H_i$ ($i = x, y, z$) marks the external magnetic field and its direction, while $J_\mathrm{c}$ is the electric current applied along the $x$-axis; $\theta$ represents the angle between saturated magnetization $M_\mathrm{s}$ and $x$-axis, while $\beta$ and $\varphi$ are angles between $H_i$ and $x$-axis.
	(b) In-plane magnetization for S4 and S7.5. The inset shows the out-of-plane magnetization. (c) Anomalous Hall resistance $R_\mathrm{AHR}$ and (d) planar Hall resistance $R_\mathrm{PHR}$ for S4 and S7.5. For $R_\mathrm{PHR}$, to fully polarize the magnetization, a field of 1\,T was applied within the film plane. The solid lines in panel (d) are fits to $R_\mathrm{PHR}(\varphi)$ = $R_\mathrm{PHR}(0)$$\sin\varphi\cos\varphi$.}
\end{figure}

In this paper, we systematically investigated the effective spin-Hall angle and the torque efficiency in perpendicularly magnetized Ta/Cu/[Ni/Co]$_{5}$/Cu-CuO$_x$ heterostructures. 
Both the SOT provided by the Ta layer and the OT, generated at the Cu-CuO$_{x}$ interface, can interact with the magnetic moments in the [Ni/Co]$_5$ multilayers. Considering that the Co and Ni have relatively large orbit-spin conversion efficiency among the 3$d$ FM metals,\cite{choi_observation_2023,lee_orbital_2021,go_theory_2020,xiao_enhancement_2022} [Ni/Co]$_5$ multilayers have been deposited as the FM layer.
The Ta(6)/Cu(3)/[Ni(0.6)/Co(0.3)]$_{5}$/Cu-CuO$_x$($t$) ($t$ = 4, 5, and 7.5) and Ta(6)/Cu(3)/[Ni(0.6)/Co(0.3)]$_{5}$/SiO$_{2}$(2) heterostructures (numbers in the bracket denote the thicknesses in nm unit) were prepared on thermally oxidized Si substrate at room temperature in an ultra-high vacuum magnetron sputtering system. 
The heterostructures with a Cu-CuO$_x$ thickness $t$ = 4, 5, and 7.5 are denoted as S4, S5, and S7.5, respectively. The Ta buffer layer was introduced to promote the (111) texture of Cu layer and thus, to induce  
a perpendicular magnetic anisotropy (PMA) in [Ni/Co]$_{5}$ FM multilayers, which was confirmed by x-ray diffraction (XRD) measurements (see Fig.~S1 and Sec. SI in the Supplementary Materials). 
The Cu-CuO$_x$ layer was produced through natural oxidation of the Cu layer in air.\cite{xiao_enhancement_2022,ding_harnessing_2020} 
The heterostructures were patterned into a Hall-bar geometry  
by the standard photolithography and Ar-ion-beam etching techniques [see Fig.~\ref{fig:fig1}(a)]. The Hall resistance and the harmonic Hall voltages were measured using a Keithley 6221 current source and a Keithley 2182 nanovoltmeter, and two Stanford S830 lock-in ampliﬁers. The magnetization was collected using a vibrating sample magnetometer (VSM, Lakeshore 7404) by applying a magnetic field both 
within and perpendicular to the film plane.

Figure~\ref{fig:fig1}(b) plots the in-plane and out-of-plane magnetization for both S4 and S7.5 heterostructures, where a clear PMA can be observed. 
The saturation field for the in-plane magnetization is $\mu_0$$H_\mathrm{s}$ = 0.72(5)\,T,
while it is only 0.052(2)\,T for the out-of-plane magnetization [see inset in Fig.~\ref{fig:fig1}(b)]. The values in the brackets denote the standard deviations.
For both S4 and S7.5 heterostructures, the same saturation magnetization $M_\mathrm{s}$ = 4.1(1)$\times$10$^{5}$~A/m was obtained, implying the 
reproducibility of FM [Ni/Co]$_5$ multilayes.
Figure~\ref{fig:fig1}(c) shows the anomalous Hall resistance (AHR) $R_\mathrm{AHR}$ versus the external magnetic field $H_z$ 
for S4 and S7.5, both exhibiting the typical features due to the presence of PMA.
The saturation $R_\mathrm{AHR}$ decreases as the thickness of Cu-CuO$_{x}$ layer increases, which is 9.6(2) and 8.4(2) m$\Omega$ for S4 and S7.5, respectively. 
Similar results have been found in Ta/Cu/[Ni/Co]$_{5}$/SiO$_{2}$ heterostructure (see Fig. S2 and Sec. SII). 
We also performed planar-Hall resistance (PHR) $R_\mathrm{PHR}$ measurements for the Ta/Cu/[Ni/Co]$_5$/Cu-CuO$_{x}$ heterostructures, which have been demonstrated to be crucial for obtaining effective spin-Hall angle.\cite{kim_layer_2013,hayashi_quantitative_2014}  Figure~\ref{fig:fig1}(d) shows the angular dependence of $R_\mathrm{PHR}(\varphi)$. 
The estimated amplitudes of $R_\mathrm{PHR}$ are 104.0(3) and 40.1(3)\,m$\Omega$ for S4 and S7.5, respectively. 

\begin{figure*}[!htp]
\centering
\includegraphics[width = 0.8\textwidth]{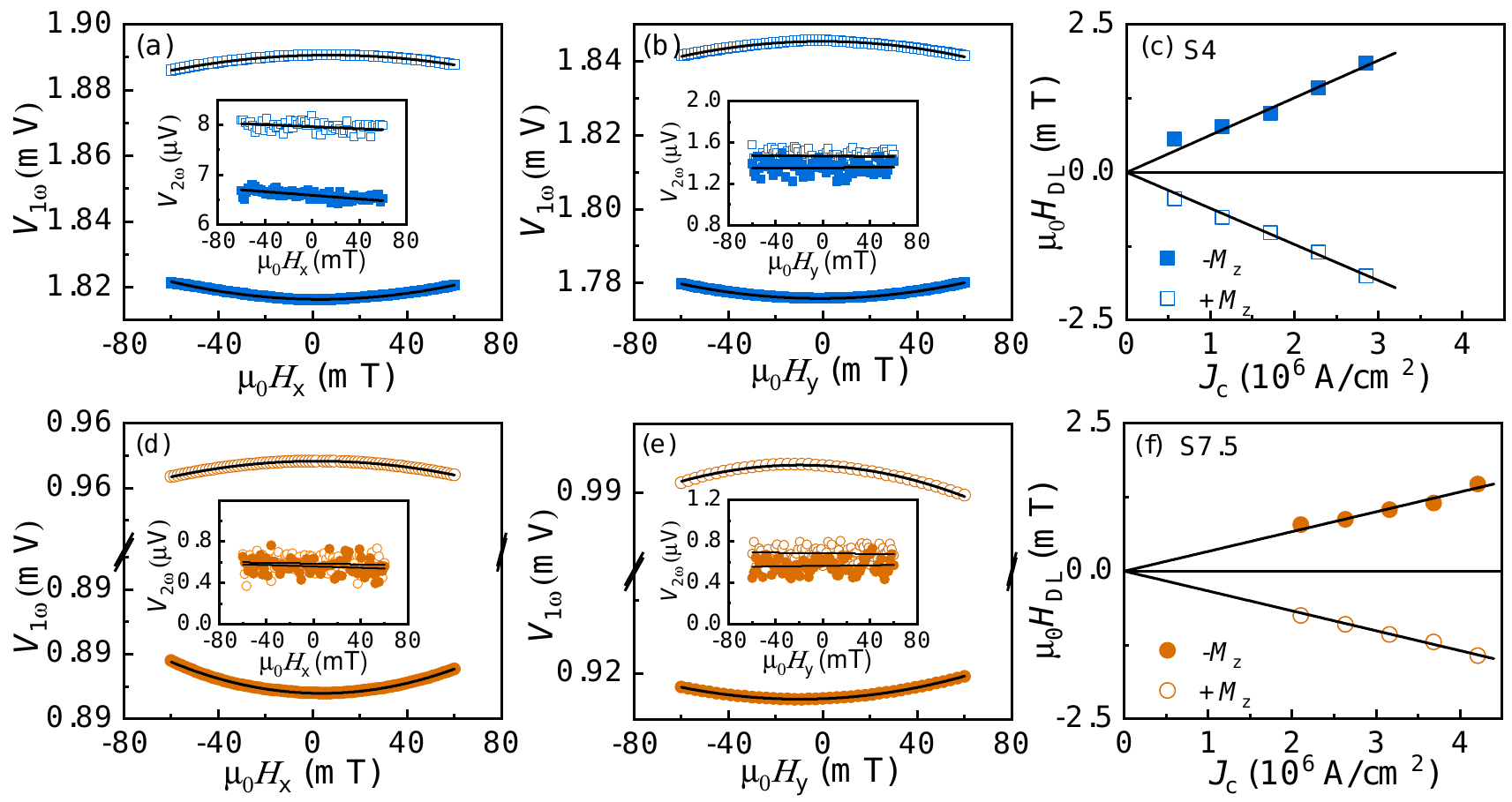}
\vspace{-3ex}
\caption{\label{fig:fig2} Harmonic Hall voltages $V_{1\omega}$ for S4 heterostructure as a function of external magnetic field applied along the $x$-axis (a) and $y$-axis (b). 
	The insets in (a) and (b) plot the field-dependent $V_{2\omega}$. (c) The SOT-induced damping-like effective field $H_\mathrm{DL}$ for S4 versus the ac electric current density $J_\mathrm{c}$. The solids lines in (a) and (b) represent the quadratic and linear fittings for $V_{1\omega}$ and $V_{2\omega}$, while the solid lines in (c) are linear fits. 
	The analog results for S7.5 heterostructure are shown in panels (d)-(f). }
\end{figure*}

In the FM/HM heterostructures, an applied electric current generates effective fields due to the presence of spin-Hall effect and the Rashba-Edelstein effect in the HM layer, which then interact with the magnetic moments in the FM layer, and eventually manipulate its magnetic states.\cite{manchon_current-induced_2019,sinova_spin_2015,liu_current-induced_2012} 
In general, such effective fields consist of two components, i.e., a damping-like field $H_\mathrm{DL}$ 
and a field-like field $H_\mathrm{FL}$. 
Both effective fields can be extracted by performing ac harmonic-Hall-voltage measurements. As shown in Fig.~\ref{fig:fig1}(a), an ac electric current was applied along the $x$-axis of the 
heterostructures, while longitudinal field $H_\mathrm{x}$ and transverse  
field $H_\mathrm{y}$ were applied within the thin-film plane. After fully polarizing the magnetization along the +$z$ and -$z$-axes, the first $V_{1\omega}$ and the second $V_{2\omega}$ harmonic Hall voltages were collected.
For S4 and S7.5, $V_{1\omega}(H)$ exhibits a parabola-like dependence for both $H_\mathrm{x}$ and $H_\mathrm{y}$, while  $V_{2\omega}(H)$ is linear in both magnetic fields [see Figs.~\ref{fig:fig2}(a)-(b) and (d)-(e)].   
Then, the effective fields $H_\mathrm{DL}$ and $H_\mathrm{FL}$ can be evaluated following $H_\mathrm{DL} = -2\frac{\partial V_{2\omega}}{\partial H_x}/\frac{\partial^2 V_{1\omega}}{\partial H_x^2}$ and $H_\mathrm{FL} = -2\frac{\partial V_{2\omega}}{\partial H_y}/\frac{\partial^2 V_{1\omega}}{\partial H_y^2}$, respectively.\cite{kim_layer_2013,hayashi_quantitative_2014}

%

The measured harmonic Hall voltage, in general, may contain the contributions from both AHR and PHR. 
Thus, the effective fields require further correction (see Sec.~SIII in the supplementary Materials).
Figures~\ref{fig:fig2}(c) and (f) plot the uncorrected $H_\mathrm{DL}$ versus the ac electric current density $J_\mathrm{c}$ for S4 and S7.5, respectively. The damping-like torque efficiency $\beta_\mathrm{DL}$ (= $H_\mathrm{DL}/J_\mathrm{c}$) is estimated to be -0.63(4)$\times$10$^{-9}$ and 0.61(2)$\times$10$^{-9}$\,Tcm$^2$A$^{-1}$ for the +$M_z$ and -$M_z$ magnetic states for S4, respectively. While for S7.5, $\beta_\mathrm{DL}$ = -0.34(1) $\times$10$^{-9}$ and 0.33(2)$\times$10$^{-9}$\,Tcm$^2$A$^{-1}$ were obtained for +$M_z$ and -$M_z$, respectively.
The effective spin-Hall angle $\theta_\mathrm{SH}^\mathrm{eff}$ 
can be calculated using:\cite{liu_current-induced_2012}
\begin{equation}
\label{eq:eq1}
\theta_\mathrm{SH}^\mathrm{eff} =  \frac{J_\mathrm{s}}{J_\mathrm{c}} = \frac{2|e| M_\mathrm{s}t_\mathrm{FM}\beta_\mathrm{DL}}{\hbar},
\end{equation}
where $J_\mathrm{s}$ is the spin current density, $e$ is the electron charge, $t_\mathrm{FM}$ is the total thickness of the FM [Ni/Co]$_{5}$ layer, and $\hbar$ is the reduced Planck constant. The estimated 
$\theta_\mathrm{SH}^\mathrm{eff}$ is  0.35(2), 0.21(3) and 0.19(1) for the S4, S5 (see
details in Fig.~S3) and S7.5, respectively.  
Since the above effective fields $H_\mathrm{DL}$ and $H_\mathrm{FL}$ are not amended, the obtained $\theta_\mathrm{SH}^\mathrm{eff}$ using Eq.~\eqref{eq:eq1} should be smaller than the intrinsic values.\cite{kim_layer_2013,hayashi_quantitative_2014}

\begin{figure}[!htp]
	\centering
	\includegraphics[width = 0.48\textwidth]{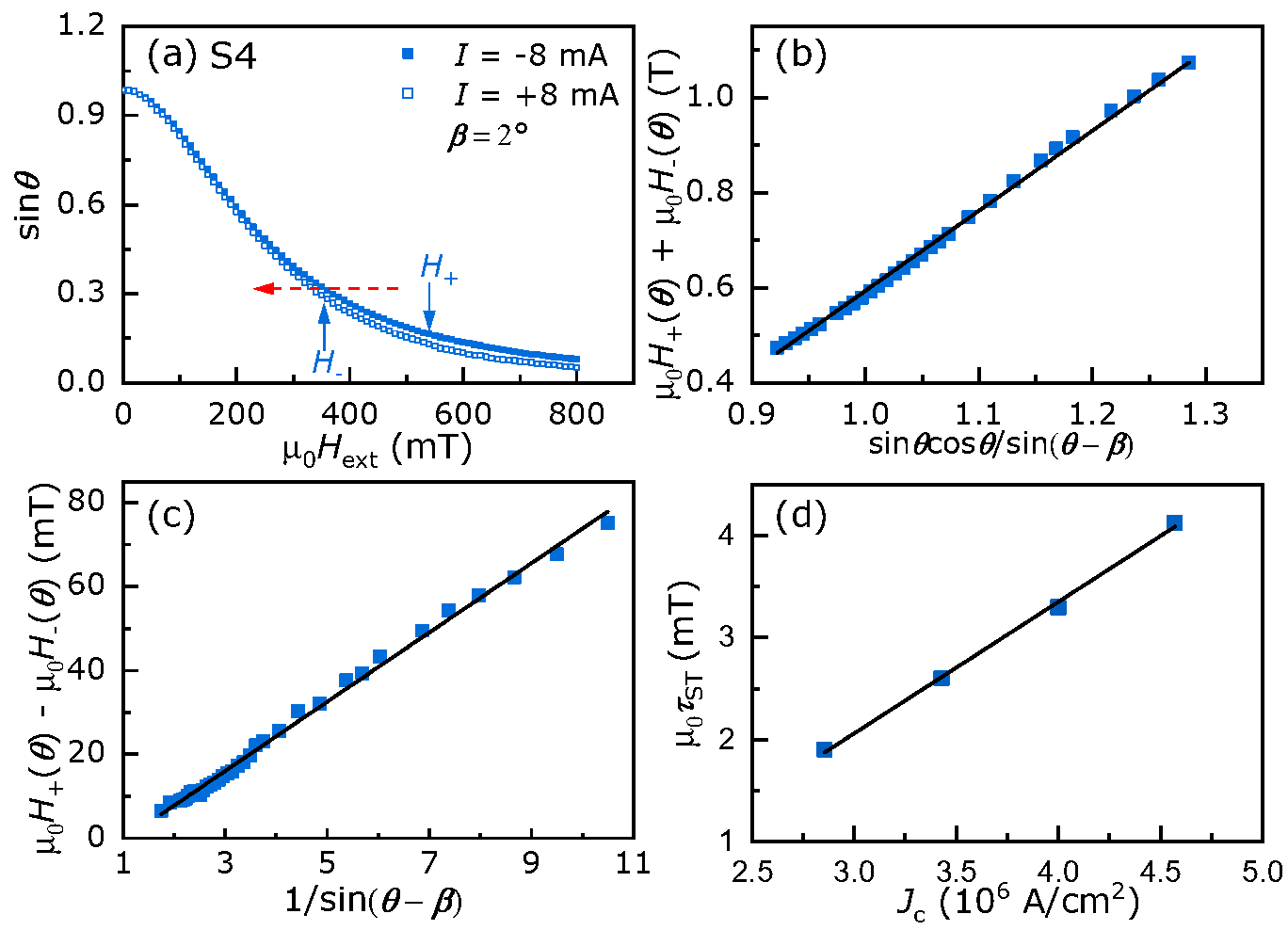}
	\vspace{-3ex}
	\caption{\label{fig:fig3} (a) The $\sin\theta$ versus $H_\mathrm{ext}$ for S4. The $H_\mathrm{ext}$ has a small angle $\beta \approx 2^\circ$ with respective to $x$-axis
		The dc current of 8\,mA was applied along both +$x$- and -$x$-axes. (b)  $H_+(\theta) + H_-(\theta)$ versus  $\sin\theta\cos\theta$/$\sin(\theta-\beta)$ and (c) $H_+(\theta) - H_-(\theta)$ versus 1/$\sin(\theta-\beta)$ 
		for S4. Solid lines are fits to Eqs. (S1) and (S2). (d) The estimated spin torque $\tau_\mathrm{ST}$ as a function of electric current density $J_\mathrm{c}$; solid line represents a linear fit.}
\end{figure} 
%

To properly extract the $\theta_\mathrm{SH}^\mathrm{eff}$ and to quantitatively measure the OT and SOT, we performed further AHR measurements based on a macro-spin model that is barely influenced by the PHR.\cite{liu_current-induced_2012,hao_giant_2015}
In this model, the spin torque $\tau_\mathrm{ST}$, the torques $\tau_\mathrm{ext}$ and $\tau_\mathrm{PMA}$ induced respectively by the external magnetic field $H_\mathrm{ext}$ and by the PMA field $H_\mathrm{PMA}$, all interact with the
magnetic moments in the FM layer.
For such AHR measurements, the magnetic field $H_\mathrm{ext}$ was applied within the $xz$-plane with a small angle $\beta \approx 2^\circ$ with respect to the $x$-axis [see details in Fig.~\ref{fig:fig1}(a)].
Such a tilted $H_\mathrm{ext}$ suppresses the formation of magnetic domains, and enables the macrospin model.\cite{liu_current-induced_2012,yun_spinorbit_2017}
Thus, the total torque $\tau_\mathrm{tot}$ that determines the magnetization rotation angle $\theta$ can be expressed as:\cite{liu_current-induced_2012} 
\begin{equation}
\label{eq:eq2}%
\begin{split}
\tau_\mathrm{tot} & = \hat{x} \cdot \left(\vec{\tau_\mathrm{ST}} + \vec{\tau_\mathrm{ext}} + \vec{\tau_\mathrm{PMA}}\right) \\
& = \tau_\mathrm{ST} + H_\mathrm{ext} \sin(\theta-\beta)- H_\mathrm{PMA} \sin\theta \cos\theta = 0.
\end{split}
\end{equation}
Here, $\tau_\mathrm{ST}= \hbar J_\mathrm{s}/2eM_\mathrm{s}t_\mathrm{FM}$, where $\hbar J_\mathrm{s}/2e$ is the spin current density.
Then, the difference of sin$\theta$ versus $H_{\mathrm{ext}}$ curves can be obtained by reversing the dc current [see Fig.~\ref{fig:fig3}(a)]. 
We define $H_+(\theta)$ and $H_-(\theta)$ as the values of $H_\mathrm{ext}$ required to produce a given value of $\theta$ when an electric current is applied along the +$x$- and -$x$-axis, respectively.
As shown in Figs.~\ref{fig:fig3}(b)-(c), both $\tau_\mathrm{ST}$ and $H_\mathrm{PMA}$ can be derived using $H_+(\theta) + H_-(\theta)$ and $H_+(\theta) - H_-(\theta)$ (see Sec.~SIV). 
Figures~\ref{fig:fig3}(d) summaries the resulting $\tau_\mathrm{ST}$ values as a function of $J_\mathrm{c}$ for S4, showing a linear feature with a slope $\tau_\mathrm{ST}$/$J_\mathrm{c}$ =  1.29(1)$\times$10$^{-9}$\,Tcm$^2$A$^{-1}$. 
By contrast, the anisotropy field $H_\mathrm{PMA}$ = 0.720(3)\,T is independent of $J_\mathrm{c}$, consistent with the magnetization measurements [see Fig.~\ref{fig:fig1}(b)].
Then, the effective spin-Hall angle can be estimated following:\cite{liu_current-induced_2012}
\begin{equation}
\label{eq:eq3}
\theta_\mathrm{SH}^\mathrm{eff} =  \frac{J_\mathrm{s}}{J_\mathrm{c}} = \frac{2|e| M_\mathrm{s}t_\mathrm{FM}}{\hbar}\cdot\frac{\tau_\mathrm{ST}}{J_\mathrm{c}}.
\end{equation}
%
For the S4, S5, and S7.5 heterostructures,  the resulting $\theta_\mathrm{SH}^\mathrm{eff}$ =  0.72(2), 0.39(1), and  0.31(1) are much higher than the values obtained using the harmonic-Hall-voltage measurement (see Fig.~\ref{fig:fig2}).
Similar behaviors have been found in other heterostructures.\cite{kim_layer_2013,yun_spinorbit_2017}

\begin{figure}[!htp]
	\centering
	\vspace{-3ex}
	\includegraphics[width=0.49\textwidth]{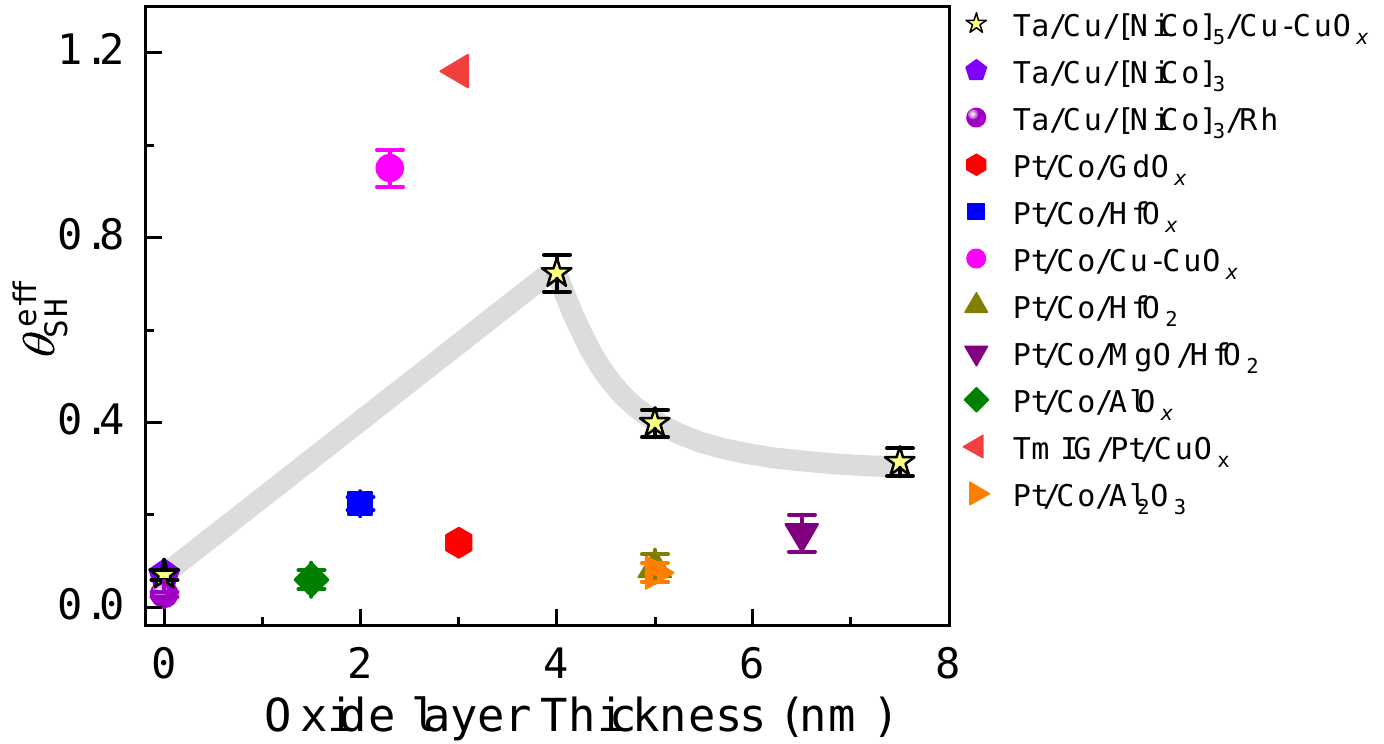}
	\caption{\label{fig:fig4}  Summary of effective spin-Hall angle $\theta_\mathrm{SH}^\mathrm{eff}$ versus the oxide layer thickness for various FM/NM or NM/FM heterostructures.
		The star symbols represent the current work on Ta/Cu/[Ni/Co]$_5$/Cu-CuO$_{x}$ heterostructures. Data of other heterostructures were taken from Ref.~\onlinecite{ding_harnessing_2020,xiao_enhancement_2022,liu_current-induced_2012,wu_enhancement_2023,hu_efficient_2022,hirai_modification_2021,cao_spin-orbit_2021}.}
	\end{figure}

The $\Delta H_\mathrm{DL}$ and thus, the $\xi$ values (the correction factor of $H_\mathrm{DL}$) for Ta/Cu/[Ni/Co]$_{5}$/Cu-CuO$_x$ heterostructures are calculated according to $\theta_\mathrm{SH}^\mathrm{eff} = \frac{2|e| M_\mathrm{s}t_\mathrm{FM}}{\hbar}\cdot\frac{\Delta H_\mathrm{DL}}{J_\mathrm{c}}$.
The resulting $\xi$ values are 0.35, 0.32 and 0.25 for S4, S5, and S7.5, respectively, which are comparable to other FM/HM heterostructures capped with an oxide layer. For instance, $\xi$ = 0.31, 0.37, and 0.61 have been found for Pt/Co/Cu-CuO$_{x}(t)$ heterostructures with $t$ = 0, 2.3, and 11.5\,nm, respectively.\cite{wu_enhancement_2023,xiao_enhancement_2022}
We now discuss why the $\xi$ values estimated from the AHR- and PHR measurements are significantly larger than the values calculated from 
$\theta_\mathrm{SH}^\mathrm{eff}$. In general, the angular-dependent planar Hall resistance follows 
$R_\mathrm{PHR}(\varphi)$ = ($R_\parallel$ - $R_\perp$)$\sin\varphi\cos\varphi$, where  
$R_\parallel$ and $R_\perp$ represent the longitudinal resistances measured with the magnetic field parallel and perpendicular to the current direction.  
Though there is no direct relationship between $R_\mathrm{PHR}$ and magnetization,\cite{seemann_origin_2011,2011Principle,mcguire_anisotropic_1975} 
$R_\mathrm{PHR}$ increases when increasing the external magnetic field, and eventually saturates once the magnetic moments are fully polarized.  
A thin Cu layer ($\leq$3\,nm) completely oxidizes into CuO$_x$ when exposing 
to air.\cite{ding_observation_2022-1,ding_harnessing_2020} While for the thick Cu layer, the oxidized CuO$_x$ layer prevents the further oxidization, and thus, it becomes the Cu-CuO$_x$ layer. 
Due to the insulating and antiferromagnetic natures,
the CuO$_x$ layer barely contributes to $R_\mathrm{AHR}$ and $R_\mathrm{PHR}$.\cite{lawrie_search_1998}
According to the previous studies, the nonmagnetic Cu layer can increase the $R_\mathrm{PHR}$ as well as enhance $H_\mathrm{PMA}$ of the FM or multilayers,\cite{bui_high-sensitivity_2013,chui_detection_2007,wang_magnetic_2013,wu_enhancement_2013,ayareh_tuning_2018} leading to a $\xi$ value being significantly larger than 1.	
It is noted that $\xi$ should be solely determined by the FM layer, the extra contribution from the Cu or CuO$_x$ layer would result in inaccurate calculations of $\Delta H_\mathrm{DL}$ and $\Delta H_\mathrm{FL}$ and thus, contribute to the intrinsic $\theta_\mathrm{SH}^\mathrm{eff}$, when analyzing the harmonic-Hall voltages.

Figure~\ref{fig:fig4} summaries the obtained $\theta_\mathrm{SH}^\mathrm{eff}$ using the AHR measurements for Ta/Cu/[Ni/Co]$_{5}$/Cu-CuO$_x$ heterostructures (see Fig.~\ref{fig:fig3}).
The $\theta_\mathrm{SH}^\mathrm{eff}$ is largely enhanced by introducing the naturally oxidized Cu-CuO$_{x}$ layer. For example, $\theta_\mathrm{SH}^\mathrm{eff}$ = 0.72 for S4 is almost one order of magnitude larger than 0.07 measured for the Cu-CuO$_x$-free Ta/Cu/[Ni/Co]$_{5}$ heterostructure (see Fig.~S4).\cite{hu_efficient_2022} We also compare  $\theta_\mathrm{SH}^\mathrm{eff}$ with other heterostructures capped with different oxide layers. Similar to our heterostructures, the FM/HM or HM/FM heterostructures capped with a Cu-CuO$_x$ layer all exhibit a very large 
$\theta_\mathrm{SH}^\mathrm{eff}$. For example, $\theta_\mathrm{SH}^\mathrm{eff}$ is 1.16 and 0.95 for the TmIG/Pt/Cu-CuO$_x$ and Pt/Co/Cu-CuO$_x$ heterostructures, respectively.\cite{ding_harnessing_2020,xiao_enhancement_2022} 
While for the Pt/Co heterostructure capped with other oxides, e.g., GdO$_x$, HfO$_x$, and AlO$_x$, the estimated $\theta_\mathrm{SH}^\mathrm{eff}$ is smaller than or comparable to the values of the 
ones without an oxide layer (see Fig.~\ref{fig:fig4}).\cite{xiao_enhancement_2022,ding_harnessing_2020,liu_current-induced_2012,wu_enhancement_2023,hirai_modification_2021}

\begin{figure}[!htp]
	\centering
	\vspace{-3ex}
	\includegraphics[width=0.48\textwidth]{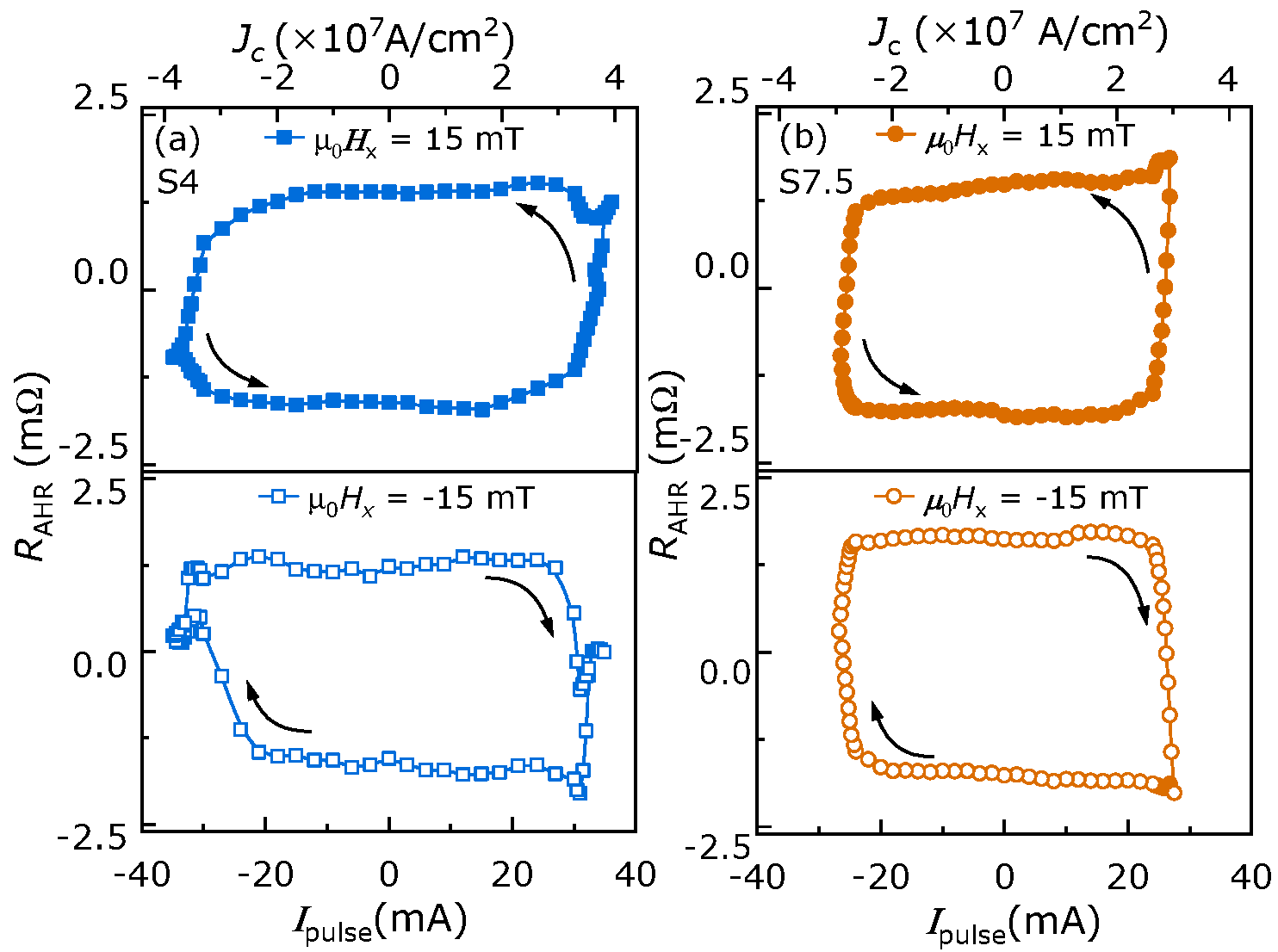}
	\caption{\label{fig:fig5} Current-induced magnetization switching. Anomalous Hall resistance $R_\mathrm{AHR}$ versus the pulsed current $I_\mathrm{pulse}$ for S4 (a) and S7.5 (b).
		For both heterostructures, an in-plane assistance magnetic field of $\mu_0$$H_x$ = $\pm$15\,mT was applied along $x$-axis [see details in Fig.~\ref{fig:fig1}(a)]. 
		The heating effects were minimized by using an ac-pulse-current method (see Sec.~SVI).
		The arrows mark the direction of current sweep.}
\end{figure}

In general, the SOT efficiency $\beta_\mathrm{DL}$ is mostly determined by the  $\theta_\mathrm{SH}^\mathrm{eff}$ [see Eq.~\eqref{eq:eq1}], the latter is usually reflected by the SOC strength of the spin-source materials. Therefore, the light elements, e.g., Cu, Al, and Mg, whose SOC is almost negligible,\cite{hirai_modification_2021,xiao_enhancement_2022} always show small $\theta_\mathrm{SH}^\mathrm{eff}$ and $\beta_\mathrm{DL}$. However, the oxidized Cu layer significantly enhances the orbital-Rashba effect (ORE), a counterpart of the spin-Rashba effect (SRE), near the Fermi level due to the strong hybridization between O-2$p$- and Cu 3$d$-orbits.\cite{go_orbital_2021}
Such ORE can be realized at the thin-film surfaces even without SOC and seems rather weak in other oxides, e.g., GdO$_x$ and HfO$_x$.\cite{go_intrinsic_2018,go_orbital_2021}
Similar to SRE that leads the spin current, ORE can generate OAM textures in the momentum space.
In the FM/Cu-CuO$_x$ heterostructures, OAM can be transferred from the Cu-CuO$_{x}$ interface to the adjacent FM layer, providing a sizable OT to switch the magnetization, analogous to the case of SOT.\cite{xiao_enhancement_2022,an_spintorque_2016,go_intrinsic_2018} The enhanced $\theta_\mathrm{SH}^\mathrm{eff}$ and $\beta_\mathrm{DL}$ are most likely due to the collaborative driven of the SOT from the HM layer and the OT at the Cu-CuO$_x$ interface.
Since the OAM accumulated at the Cu-CuO$_x$ interface decays rapidly, to inject the OAM into the adjacent FM layer and to produce a sizable OT, the thickness of Cu layer is limited.\cite{xiao_enhancement_2022,go_orbital_2021}  
An ultrathin Cu layer (e.g., $t_\mathrm{Cu}$ $\leq$ 3\,nm) can be entirely oxidized,\cite{ding_observation_2022-1} consequently, the OAM can reach the adjacent FM layer without a clear relaxation. 
While for the thicker Cu layer, the OAM rapidly decays in the Cu layer before reaching the adjacent FM layer due to its nonconserved nature.\cite{go_orbital_2021} The estimated critical Cu-layer thickness is about 20\,nm (see Fig.~S5).
Therefore, instead of  continuously decreasing, the $\theta_\mathrm{SH}^\mathrm{eff}$ starts to saturate when increasing the $t_\mathrm{Cu}$ up to the threshold value.

Since the torque efficiency is largely enhanced by introducing a naturally oxidized Cu-CuO$_x$ layer, we also checked the current-induced magnetization switching in S4 and S7.5 heterostructures. 
As shown in Fig.~\ref{fig:fig5}, for both heterostructures, $R_\mathrm{AHR}$ as a function of the pulse $I_\mathrm{pulse}$ shows anticlockwise and clockwise loops by applying an assistance magnetic field $\mu_0$$H_x$ = -15 and +15\,mT. The critical currents for switching the magnetization are identified as $I_\mathrm{c}$ = 28.1 and 26.3\,mA for S4 and S7.5, respectively. Since the resistivity of the metallic layers is of the same order of magnitude, we assume that the current uniformly flowing through the metallic layers,\cite{xiao_enhancement_2022} the estimated critical current density $J_\mathrm{c}$ are 3.21$\times$10$^7$ and 2.94$\times$10$^7$\,A/cm$^2$ for S4 and S7.5, which are comparable to other HM/FM/Cu-CuO$_x$ heterostructures,\cite{xiao_enhancement_2022} but are clearly lower than the Cu-CuO$_x$ free heterostructures (see Fig.~S6).\cite{p_spin_2017,xie_situ_2020,cao_spin-orbit_2021}
The reduced $J_\mathrm{c}$ is also reflected by the enhanced $\theta_\mathrm{SH}^\mathrm{eff}$ and $\beta_\mathrm{DL}$ in Ta/Cu/[Ni/Co]$_5$/Cu-CuO$_{x}$ heterostructures. 
Though S4 has a larger $\theta_\mathrm{SH}^\mathrm{eff}$ than S7.5, both heterostructures exhibit comparable critical current densities (see Fig.~\ref{fig:fig5}). 
The uncertainty of CuO$_x$-layer thickness and the different valence states of Cu ions might prevent properly estimating the current density in the Ta/Cu/[Ni/Co]$_{5}$/Cu–CuO$_{x}$, both require further experimental investigations.

In summary, we experimentally observed a signiﬁcant increase in the torque efﬁciency,  $\theta_\mathrm{SH}^\mathrm{eff}$, and $\beta_\mathrm{DL}$ in the perpendicularly magnetized HM/FM/NM multilayers through surface oxidation. In addition, we also observed a remarkable contribution of surface-oxidized light metal to the PHR under a large magnetic field, which affects the correction of $\Delta H_{\mathrm{DL}(\mathrm{FL})}$  and $\theta_\mathrm{SH}^\mathrm{eff}$. 
Our results suggest that the large enhancement of $\theta_\mathrm{SH}^\mathrm{eff}$ originates from the OT derived from the OAM at the Cu–CuO$_{x}$ interface and the SOT provided by the Ta layer. The 
estimated $\theta_\mathrm{SH}^\mathrm{eff}$ = 0.72(2) for Ta/Cu/[Ni/Co]$_{5}$/Cu–CuO$_{x}$(4) is obviously lager than other well-studied 5$d$ metals.
Our results suggest that the combination of SOT and OT represents one of the most efficient method to enhance the $\theta_\mathrm{SH}^\mathrm{eff}$ and $\beta_\mathrm{DL}$ , and thus, to improve the torque efficiency in the low power consumption spintronic devices.

\vspace{3mm}
\noindent\textbf{Supplementary Material}\\
See the supplementary material for x-ray diffraction measurements, ac harmonic-Hall-voltage measurements on S5 heterostructure, and estimation of the critical thickness of Cu layer.
It also contains magnetic characterizations, dc Hall and magnetization switching measurements on the Cu-CuO$_x$ free Ta/Cu/[Ni/Co]$_5$/SiO$_2$ heterostructure.


%
%

%

\begin{acknowledgments}
This work was supported by  
the Natural Science Foundation of Shanghai 
(Grants No.\ 21ZR1420500 and 21JC\-140\-2300), Natural Science
Foundation of Chongqing (Grant No.\ 2022NSCQ-MSX1468), the National Natural Science Foundation of China (No. 
12174103 and No. 12374105). Y.X.\ acknowledges support from the Shanghai Pujiang Program (Grant No.\ 21PJ1403100).
\end{acknowledgments}

\section*{AUTHOR DECLARATIONS}
\subsection*{Conflict of Interest}
The authors have no conflicts to disclose.
\subsection*{Author Contributions}

\textbf{Kun Zheng}: Data curation (lead); Formal analysis (equal); Writing – original draft (lead). \textbf{Cuimei Cao}: Data curation (equal); Investigation (equal). \textbf{Yingying Lu}: Data curation (equal). \textbf{Jing Meng}: Formal analysis (equal). \textbf{Junpeng Pan}: Supervision (equal). \textbf{Zhenjie Zhao}: Data curation (supporting); Methodology (equal). \textbf{Yang Xu}: Supervision (equal); Writing – review and editing (supporting). \textbf{Tian Shang}: Supervision (lead); Writing – original draft (equal); Writing – review and editing (equal). \textbf{Qingfeng Zhan}: Conceptualization (equal); Supervision (equal); Writing – review and editing (equal).

\section*{DATA AVAILABILITY}
The data that support the findings of this study are available from the corresponding author upon reasonable request.

\bibliography{TaCuNiCo}

\end{document}